\documentstyle[mbf,wrapft]{ptptex}

\def\bq{{\mbf q}}
\def\bk{{\mbf k}}

\def\SM3{\Sigma N (3/2)}
\def\SN1{\Sigma N (1/2)}

\def\TS1{\hbox{}^3S_1}
\def\TD1{\hbox{}^3D_1}

\def\eq#1{Eq.~(\ref{#1})}

\preprintnumber{
KUNS-1752 \\
nucl-th/0201052 \\ January 2002}
\title{Redundant Components in the $3\alpha$ Faddeev Equation \\
Using $2\alpha$ RGM Kernel}

\author{
Yoshikazu {\sc Fujiwara}, Yasuyuki {\sc Suzuki}$^{*}$,
Kazuya {\sc Miyagawa}$^{**}$ \\
Michio {\sc Kohno}$^{***}$,
Hidekatsu {\sc Nemura}$^{****}$ \\
}

\inst{
Department of Physics, Kyoto University,
Kyoto 606-8502, Japan \\
$\hbox{}^{*}$Department of Physics, Niigata University,
Niigata 950-2181, Japan \\
$\hbox{}^{**}$Department of Applied Physics, Okayama Science University,
Okayama 700-0005, Japan \\
$\hbox{}^{***}$Physics Division, Kyushu Dental College,
Kitakyushu 803-8580, Japan \\
$\hbox{}^{****}$Institute of Particle and Nuclear Physics, KEK,
Ibaragi 305-0801, Japan
}
\recdate{
\today
}
\abst{
The $3\alpha$ Faddeev equation using $2\alpha$ RGM kernel involves 
redundant components whose contribution to the total wave function
completely cancels out.
We propose a practical method to solve this Faddeev equation,
by eliminating the admixture of such redundant components.
A complete equivalence between the present Faddeev approach and
a variational approach using the translationally invariant
harmonic-oscillator basis is numerically shown
with respect to the $3\alpha$ bound state corresponding
to the ground state of $\hbox{}^{12}\hbox{C}$.
}

\begin{document}

\maketitle

A first issue for applying realistic quark-model
baryon-baryon interactions to few-baryon systems,
such as the hypertriton, is to find a basic
three-cluster equation which is formulated by using
microscopic two-cluster quark-exchange kernel
of the resonating-group method (RGM).
This issue is non-trivial, not only because the quark-exchange kernel
is non-local and energy dependent, but also because RGM equations
sometimes involve redundant components due to the effect of
the antisymmetrization; i.e., the Pauli-forbidden states. 
A desirable feature of such a three-cluster equation is that
it can be solved in either or both of the variational approach
and the Faddeev formalism, yielding completely the same result.
In our previous paper \cite{FU01}, which is referred to I hereafter,
we have proposed a simple three-cluster equation, which is similar to the
orthogonality condition model (OCM), \cite{HO74} but employs
the two-cluster RGM kernel as the interaction potential.
The three-cluster Pauli-allowed space is constructed
by the orthogonality of the total wave functions
to the pairwise Pauli forbidden states. Although this definition
of the three-cluster Pauli-allowed space is not exactly equivalent to
the standard definition given by the three-cluster normalization kernel,
this assumption is essential to find a complete equivalence between the
proposed three-cluster equation and the Faddeev equation which
employs a singularity-free $T$-matrix derived
from the RGM kernel (the RGM $T$-matrix). \cite{FU01}

More explicitly, the Faddeev equation for systems composed
of three identical bosons is expressed as
\begin{eqnarray}
\lambda \varphi=G_0 \widetilde{T} S \varphi\ ,
\label{eq1}
\end{eqnarray}
where $S$ implies the real symmetric ($\hbox{}^tS=S$) matrix
for rearrangement of the three-types of the Jacobi-coordinates:
$\varphi_\beta+\varphi_\gamma=[(123)+(123)^2]\,\varphi_\alpha
=S \varphi_\alpha$. Note that $1+S$ is semi-positive definite.
In \eq{eq1}, $\varphi=\varphi_\alpha$ is the $\alpha$-component of the
total wave function $\Psi=\varphi_\alpha+\varphi_\beta+\varphi_\gamma
=(1+S)\varphi$, $G_0=G_0(E)=1/(E-H_0+i0)$ is the 3-body
free Green function for the negative energy $E$,
and $\widetilde{T}=\widetilde{T}^{(3)}
(E, \varepsilon)$ is essentially the two-cluster $T$-matrix
derived from the RGM kernel (see Eqs.\,(2.8), (2.9) and (3.20) of I).
This  $\widetilde{T}$ satisfies the basic relationship
(Eq.\,(2.23) of I)
\begin{eqnarray}
\langle u| [1+G_0 \widetilde{T}]=[1+ \widetilde{T}G_0] |u \rangle=0\ .
\label{eq2}
\end{eqnarray}
Here we assume only one Pauli forbidden state $|u\rangle$,
for simplicity.

In order to find a trivial solution of \eq{eq1}, we first solve
\begin{eqnarray}
S|uf^\tau \rangle=\tau |uf^\tau \rangle\ ,
\label{eq3}
\end{eqnarray}
where $|uf\rangle=|u\rangle |f\rangle$ is a product of
two functions corresponding to the two momentum Jacobi-coordinate
vectors, $\bk$ and $\bq$, respectively.
Each function of $|u\rangle$ and $|f\rangle$ is assumed
to be normalized as $\langle u|u\rangle=\langle f|f\rangle=1$.
In the $3d^\prime$ system discussed in I, there appears
no $\tau=-1$ eigen-value, while in the $3\alpha$ system we find
two solutions with $\tau=-1$, as is shown below.
We find these by solving the eigen-value problem
\begin{eqnarray}
\langle u|S|uf^\tau \rangle=\tau |f^\tau \rangle
\label{eq4}
\end{eqnarray}
with smaller dimensionality than \eq{eq3}. In fact,
if we have a $\tau=-1$ solution $|f^\tau \rangle$ for \eq{eq4},
the relationship $(1+S)^2=[1+(123)+(123)^2]^2=3(1+S)$ implies
that $\langle u|(1+S)^2|uf^\tau\rangle=0$. If we further
use $S^\dagger=S$, this leads to $\langle (1+S)uf^\tau
|(1+S)uf^\tau \rangle=0$, namely,
$(1+S)|uf^\tau \rangle=0$ for $\tau=-1$.
Such a solution with $\tau=-1$ corresponds to the [21] symmetric 
component of the total wave function with respect to the
permutations of the three bosons.
Let us assume that we only have one $\tau=-1$ solution
for simplicity of discussion.
The Faddeev component constructed from this solution
\begin{eqnarray}
\varphi^\tau_0=G_0 |uf^\tau \rangle \qquad \hbox{with} \quad \tau=-1
\label{eq5}
\end{eqnarray}
gives vanishing contribution to the total wave function $\Psi$,
since $G_0$ is [3] symmetric. 
On the other hand, it is very easy to show that $\varphi^\tau_0$ is
a $\lambda=1$ solution of \eq{eq1}:
\begin{eqnarray}
\varphi^\tau_0=G_0 \widetilde{T} S \varphi^\tau_0\
\qquad \hbox{for} \quad \tau=-1\ .
\label{eq6}
\end{eqnarray}
This is because of the very special nature of $\widetilde{T}$,
satisfying the basic relationship \eq{eq2}.

Since our true solution of \eq{eq1} with $\lambda=-1$ includes
such a component as $\varphi^\tau_0$, a general solution of
\eq{eq1} with $\lambda=-1$ is expressed as
\begin{eqnarray}
\varphi=\widetilde{\varphi}+C\varphi^\tau_0\ ,
\label{eq7}
\end{eqnarray}
where $\widetilde{\varphi}$ is a special solution
of \eq{eq1} with $\lambda=-1$. An optimum choice of the
coefficient $C$ in \eq{eq7} is to make the value $\langle \varphi|
1+S|\varphi\rangle/\langle \varphi|\varphi \rangle$ maximum.
If we expand $\varphi$ by the complete orthonormalized basis
of \eq{eq3} (including the $\tau=-1$ solutions), we find that
this condition is expressed as the orthogonality
\begin{eqnarray}
\langle uf^\tau|\varphi \rangle=0 \qquad \hbox{for} \quad \tau=-1\ .
\label{eq8}
\end{eqnarray}
Namely, we only need to solve \eq{eq1} under the condition
of \eq{eq8}.

The following procedure to eliminate the redundant
components with $\tau=-1$ from the Faddeev equation is
very similar to the technique used for finding a unique solution
of RGM equations. We modify \eq{eq1} as
\begin{eqnarray}
\lambda \varphi=\left[ G_0 \widetilde{T} S -G_0|uf^\tau\rangle
{1 \over \langle uf^\tau|G_0|uf^\tau\rangle}
\langle uf^\tau|\,\right]\,\varphi\ .
\label{eq9}
\end{eqnarray}
If we multiply $\langle uf^\tau|$ from the left,
$\langle uf^\tau|G_0 \widetilde{T} S=\langle uf^\tau|$ yields
\begin{eqnarray}
\lambda \langle uf^\tau|\varphi \rangle=0\ .
\label{eq10}
\end{eqnarray}
Therefor, the solution $\varphi$ of \eq{eq9} satisfies \eq{eq8} so
long as $\lambda \neq 0$.
The fact that $\varphi^\tau_0$ is a $\lambda=0$ solution
of \eq{eq9} implies that the trivial $\lambda=1$ solution
of \eq{eq1} is pushed down to the $\lambda=0$ eigen-state in \eq{eq9}.
The solution of \eq{eq9} also satisfies \eq{eq1} when $\lambda \neq 0$,
since the second term of \eq{eq9} does not contribute because
of \eq{eq8}. If we multiply \eq{eq9} only with $\langle u|$ from
the left and leave the degree of freedom for $\bq$, we obtain
\begin{eqnarray}
\lambda \langle u|\varphi\rangle=-\langle u|S\varphi\rangle
-\langle u|G_0|uf^\tau\rangle
{1 \over \langle uf^\tau|G_0|uf^\tau\rangle}
\langle uf^\tau|\varphi\rangle\ .
\label{eq11}
\end{eqnarray}
Again, the second term of the right hand side does not contribute
if $\lambda \neq 0$, and we find
\begin{eqnarray}
\langle u|\lambda +S|\varphi\rangle=0
\qquad \hbox{for} \quad \lambda \neq 0\ .
\label{eq12}
\end{eqnarray}
In particular, the $\lambda=1$ solution of \eq{eq9} leads
to the condition
\begin{eqnarray}
\langle u|1+S|\varphi\rangle=0
\longrightarrow \langle u|\varphi_\alpha+\varphi_\beta
+\varphi_\gamma\rangle=0\ ,
\label{eq13}
\end{eqnarray}
which implies that our total wave function $\Psi$ does not
contain the pairwise redundant component $\langle u|\Psi\rangle=0$.
In summary, solving \eq{eq9} automatically guarantees the solution
\begin{eqnarray}
\lambda \varphi=G_0 \widetilde{T} S \varphi\ ,\quad
\langle uf^\tau|\varphi \rangle=0\ ,\quad
\langle u|\lambda+S|\varphi\rangle=0 \qquad \hbox{for} \quad 
\lambda \neq 0\ .
\label{eq14}
\end{eqnarray}

\begin{table}[b]
\caption{$|uf\rangle$ $SU_3$ states for 3 $\alpha$ system.
$N=N_1+N_2$ is the total h.o. quanta of the whole system.
Only non-negative $\lambda$ and $\mu$ are allowed.
}
\label{table1}
\begin{center}
\renewcommand{\arraystretch}{1.1}
\setlength{\tabcolsep}{4mm}
\begin{tabular}{ccc}
\hline
$N_1$ & $N_2$ & $(\lambda \mu)$ \\
\hline
0  & $N$ &   $(N 0)$   \\
2  & $N-2$ & $(N 0)$, $(N-2, 1)$, $(N-4, 2)$  \\
\hline
\end{tabular}
\end{center}
\end{table}

Next, let us consider the analytic solution of \eq{eq4} in
a particular case of the $3\alpha$ system. It is convenient
to use the translationally invariant harmonic-oscillator (h.o.) basis
used in I. The Pauli forbidden states of the $2\alpha$ system
consist of the h.o. states with $N_1=0$ (for the relative
angular momentum $\ell=0$) and $N_1=2$ (for $\ell=0$ and 2),
where $N_1$ is the total h.o. quanta ($2n+1$) for the variable $\bk$.
The odd $N_1$ state is omitted from the very beginning, by assuming
only even partial waves for the $2\alpha$ relative motion.
We have to solve the problem which $SU_3$ states are classified
to the [21] symmetry among the $SU_3$-coupled 2-particle h.o. states
$[U_{(N_1 0)}(\bk) U_{(N_2 0)}(\bq)]_{(\lambda \mu)a}$.
For $N_1=0$ and 2, these $SU_3$ states are constructed as
in Table \ref{table1}.
Since we know that all the $SU_3$ state of the $3\alpha$ system
is Pauli forbidden for the total h.o. quanta $N=N_1+N_2 \leq 6$,
it is sufficient to consider only the states given in Table \ref{table2}.
\begin{table}[t]
\caption{$[3]$ and [21] symmetric basis for $|uf\rangle$,
classified by the $SU_3$ basis with $(\lambda \mu)$.}
\label{table2}
\begin{center}
\renewcommand{\arraystretch}{1.1}
\setlength{\tabcolsep}{4mm}
\begin{tabular}{cccc|c|c}
\hline
$N$ & $N_1$ & $N_2$ & $(\lambda \mu)$ & [3] & [21] \\
\hline
0 & 0 & 0 &   (0 0) & (00) & $-$  \\
2 & 0 & 2 &   (2 0) &      &      \\
  & 2 & 0 &   (2 0) & (20) & (20) \\
4 & 0 & 4 &   (4 0) &      &     \\
  & 2 & 2 &   (4 0), (21), (02) & (40), (02) & (40), (21) \\
6 & 0 & 6 &   (6 0) &      &     \\
  & 2 & 4 &   (6 0), (41), (22) & $(60)^2$, (22) & (41) \\
\hline
\end{tabular}
\end{center}
\end{table}
This result is obtained by enumerating all the [3] symmetric basis
states using the Moshinsky's method \cite{MO96}.\footnote{Here, again,
we can explicitly construct the basis states using the theory
of Double Gel'fand polynomials. \cite{KI83}.}
For example, for the two independent (20) states constructed 
from $N_1=2,~N_2=0$ and $N_1=0,~N_2=2$ h.o. bases,
only one (20) state is [3] symmetric, while the other (20) state
belongs to the [21] symmetry.\footnote{A simple discussion using
the tensor components of the 3-dimensional Jacobi-coordinate
vectors $\bk$ and $\bq$ also leads to this conclusion.
Namely, the 21-dimensional representation of the $N=2$ 2-particle
h.o. states, $\bk_\alpha \bk_\beta$ (6 dim.),
$\bq_\alpha \bq_\beta$ (6 dim.),
$\bk_\alpha \bq_\beta$ (9 dim.), is decomposed
into the following $SU_3$ irreducible representations;
[3](20): $\bk_\alpha \bk_\beta+\bq_\alpha \bq_\beta$ (6 dim.),
[21](20): $\bk_\alpha \bk_\beta-\bq_\alpha \bq_\beta$ (6 dim.),
[21](20): $\bk_\alpha \bq_\beta+\bq_\alpha \bk_\beta$ (6 dim.),
[111](01): $\epsilon_{\alpha \beta \gamma}\bk_\beta \bq_\gamma$
(3 dim.). The [21](20) state in \protect\eq{eq15} corresponds
to the second $SU_3$ state, which is [2] symmetric with respect to the
exchange of the first two particles, but is not totally [3] symmetric.}
Similarly, only one [3] symmetric (40) state is made for $N=4$,
and another [21] symmetric Pauli forbidden state is constructed
for the total angular-momentum $L=0$ states of the $3\alpha$ system.
These [21] symmetric $SU_3$ states are explicitly given by
\begin{eqnarray}
& & \varphi^{[21](20)}_a = {1 \over \sqrt{2}}
\left[\,U_{(20)a}(\bq)-U_{(20)a}(\bk)\,\right]\ \ , \nonumber \\
& & \varphi^{[21](40)}_a = \sqrt{{2 \over 5}}
U_{(40)a}(\bq)-\sqrt{{3 \over 5}}
\left[ U_{(20)}(\bk)U_{(20)}(\bq) \right]_{(40)a}\ .
\label{eq15}
\end{eqnarray}
Note that the (21) and (41) $SU_3$ states are not possible
for $L=0$. This analysis shows that the two $\tau=-1$ solutions
of \eq{eq4} for the $L=0$ $3\alpha$ system is nothing
but the [21](20) and [21](40) $SU_3$ states in \eq{eq15}.
It is also apparent why we have no $\tau=-1$ solution for
the $3d^\prime$ system. In this system, the Pauli forbidden state
is only $(0s)$ state for $2d^\prime$, and there exists no [21] symmetric
Pauli forbidden state constructed for $N=0$.

It should be noted that the existence of the [21] symmetric trivial
solutions in the original Faddeev equation (\ref{eq1}) is essential
to eliminate the three-cluster Pauli forbidden states, which can not be
trivially eliminated only through the pairwise orthogonality conditions
with respect to the variable $\bk$.
The mechanism to guarantee such a favorable result is furnished
by the cooperative role with the exchange symmetry of the boson
system and the partial elimination of the functional space $f(\bq)$
corresponding to the other variable $\bq$.

It would be legitimate to ask why such [21] symmetric components
admixed to the Faddeev component $\varphi_\alpha$ do not
play an important role in the usual Faddeev equations.
Suppose we have a situation $\varphi_\beta+\varphi_\gamma
=S \varphi_\alpha=-\varphi_\alpha$.
Then the usual Faddeev equation,
$\varphi_\alpha=G_0 T_\alpha (\varphi_\beta+\varphi_\gamma)$,
becomes $\varphi_\alpha=-G_0 T_\alpha \varphi_\alpha$.
If one uses the relationship $G_0 T_\alpha=G_\alpha V_\alpha$,
this equation is reduced to $(E-H_0)\varphi_\alpha=0$,
and $\varphi_\alpha$ turns out to be the plane wave.
This implies that we have no square-integrable trivial solutions.
Even if we use $\lambda \varphi_\alpha
=G_0 T_\alpha (\varphi_\beta+\varphi_\gamma)$, we obtain
\begin{eqnarray}
\left[\,E-H_0-\left(1-{1 \over \lambda}\right)V_\alpha \,\right]=0\ ,
\label{eq16}
\end{eqnarray}
which implies that there are no square-integrable solutions
below the 2-body threshold. (Note that we are interested 
in the situation $\lambda>0$.)
Since the [21] components admixed to $\varphi_\alpha$ do not
contribute to $\Psi$ anyway, there is no need to worry about
such admixture. 

For a practical application of the basic equation \eq{eq9} to
the $3\alpha$ system, the self-consistency procedure for determining
the energy dependence of the exchange term $\varepsilon K$ in the
allowed space, discussed in I, is very important.
We therefore need to evaluate the expectation value $\varepsilon$ of
the $2\alpha$ Hamiltonian through
\begin{eqnarray}
\varepsilon={1 \over 3}E+{1 \over 2} \langle \varphi |
H_0 (1+S)|\varphi \rangle\ ,
\label{eq17}
\end{eqnarray}
where $H_0$ is the 3-body free kinetic-energy operator
and the Faddeev component $\varphi$ is normalized
as $3\langle \varphi |1+S| \varphi \rangle=1$.
Starting from some specific values of $\varepsilon$ and $E$,
we solve \eq{eq9} and find a negative 3-body energy $E$ such that
the eigen-value $\lambda(E)$ becomes 1.
The normalized Faddeev component $\varphi$ yields
a new vale of $\varepsilon$ through \eq{eq17}.
Since it is usually not equal to the starting value,
we repeat the process by using the new value.
This process of double iteration converges very fast if
the starting values of $\varepsilon$ and $E$ are properly chosen.
For numerical calculation, we discretize the continuous
variables $k$ and $q$, using the Gauss-Legendre $n_1$- and $n_2$-point
quadrature formula, respectively,
for each of the three intervals of 0 - 1 - 3 - 6 $\hbox{fm}^{-1}$. 
The small contribution from the intermediate integral
over $k$ beyond $k_0=6~\hbox{fm}^{-1}$ in the $2\alpha$ $T$-matrix
calculation is also taken into account by using
the Gauss-Legendre $n_3$-point quadrature formula through the
mapping $k=k_0+{\rm tan}(\pi/4)(1+x)$.\footnote{These $n_3$ points
for $k$ are not included for solving the Faddeev
equation (\protect\ref{eq9}),
since it causes a numerical inaccuracy for the interpolation.}
The momentum region $q=$ 6 $\hbox{fm}^{-1}$ - $\infty$ is
also discretized by the $n_3$ point formula just
as in the $k$ discretization case. 
The values of $n_1$-$n_2$-$n_3$ are given in Table \ref{table3}.
The partial-wave decomposition of
the $2\alpha$ RGM kernel is carried out numerically using
the Gauss-Legendre 20-point quadrature formula.
The incorporation of the Coulomb force to the Faddeev formalism
is only possible by using the shielded Coulomb
potential $u(r)=(1/r)\theta(R_C-r)$,
where $\theta(x)$ is the Heaviside step function.
Here we use a rather small value $R_C=6~\hbox{fm}$, in order
to make the phase-shift calculation of the $2\alpha$ system
for higher partial-waves numerically stable.
The Coulomb exchange kernel for this interaction is explicitly calculated.
Since our Faddeev component $\varphi$ oscillates due to the 
orthogonality to the redundant components, we need to include
the $2\alpha$ partial waves at least up to $\ell=4$. The convergence
of the result is confirmed by including the partial waves up to $\ell \leq 8$.
The numerical inaccuracy of the angular-momentum projection is
examined by extending the 20-point quadrature formula
to the 30-point formula, and we found that the error is less than 1 keV
if numbers of discretization points for $k$ and $q$ are large enough.
Furthermore, the modified spline interpolation technique
developed by Gl{\" o}ckle et al. \cite{GL82} is employed for
constructing the rearrangement matrix $S$.
For the diagonalization of the large non-symmetric matrix,
the Arnordi-Lanczos algorithm recently developed in the ARPACK
subroutine package \cite{AR96} is very useful.
For the effective 2-nucleon force for the $2\alpha$ RGM kernel,
we use the Volkov No.\,2 force with $m=0.59$,
following the $3\alpha$ RGM calculation by Fukushima
and Kamimura \cite{FU78}.
The h.o. constant for the $\alpha$ cluster is assumed
to be $\nu=0.275~\hbox{fm}^{-2}$.

\begin{table}[b]
\caption{$3\alpha$ Faddeev calculation for $L=0$,
without the Coulomb force (with the shielded Coulomb force
with $R_C=6~{\rm fm}^{-1}$).
The discretization points of $k$ and $q$ are specified
by $n_1$-$n_2$-$n_3$ (see the text), 
$\ell_{\rm max}$ implies the maximum partial waves included,
and $n_{\rm max}=(\ell/2+1)(3n_1)(3n_2+n_3)$ the dimension
of the diagonalization for the Faddeev equation (\protect\ref{eq9}).
The effective 2-nucleon force for the $2\alpha$ RGM kernel
is Volkov No. 2 with $m=0.59$.} 
\label{table3}
\begin{center}
\renewcommand{\arraystretch}{1.5}
\setlength{\tabcolsep}{2mm}
\begin{tabular}{cccccccc}
\hline
$\ell_{\rm max}$ & $n_1$-$n_2$-$n_3$ & $n_{\rm max}$ &
$\epsilon(2\alpha)$ & $E(3\alpha)$ & $c_{(04)}$ \\
\hline
4 & 15-10-5  & 4,725 & 8.5826 (9.4542) & $-11.2023$ ($-5.9313$)
& 0.8262 (0.7905) \\
  & 20-10-10 & 7,200 & 8.5825 (9.4541) & $-11.2032$ ($-5.9318$)
& 0.8262 (0.7905) \\
\hline
6 & 15-10-5  &  6,300 & 8.4490 (9.3511) & $-11.4151$ ($-6.1188$)
& 0.8212 (0.7861) \\
  & 20-10-10 &  9,600 & 8.4488 (9.3509) & $-11.4158$ ($-6.1193$)
& 0.8212 (0.7861) \\
\hline
8 & 15-10-5  &  7,875 & 8.4467 (9.3492) & $-11.4179$ ($-6.1212$)
& 0.8211 (0.7860) \\
  & 20-10-10 & 12,000 & 8.4465 (9.3490) & $-11.4187$ ($-6.1217$)
& 0.8211 (0.7860) \\
\hline
\end{tabular}
\end{center}
\end{table}

Table \ref{table3} shows the solution of the Faddeev equation 
for the $3\alpha$ system, obtained by solving \eq{eq9}.
We find that the convergence is satisfactory,
if the relative $2\alpha$ partial waves
up to $\ell_{\rm max}=8$ are taken into account.
We compare the best result of the Faddeev calculation
with the variational calculation employing the [3] symmetric
translationally-invariant h.o. basis. The model is explained in I.
The numbers without the parentheses in Table \ref{table4} indicate
the results without the Coulomb force, while those in the parentheses
when the shielded Coulomb force with $R_C=6~{\rm fm}^{-1}$ is included.
(Note that the $3\alpha$ energy with the present shielded Coulomb
force is slightly (about 150 keV) more attractive in comparison with the
that of the full Coulomb calculation in Table I of I.)
Table \ref{table4} also lists the expectation value
of the $2\alpha$ subsystem, $\varepsilon$, and the overlap with
the simple shell-model wave function $c_{(04)}$ as well.
Here the $SU_3$ (04) wave function is expressed in the $3\alpha$ cluster
model as
\begin{eqnarray}
& & \varphi^{[3](04)}_a=\left[ U_{(40)}(\bk)U_{(40)}(\bq) \right]_{(04)a}
={8 \over 15}R_{20}(k, b_1) R_{20}(q, b_2)
Y_{(00)0}(\widehat{\bk},\widehat{\bq}) \nonumber \\
& & - {4 \over 3\sqrt{5}}R_{12}(k, b_1) R_{12}(q, b_2)
Y_{(22)0}(\widehat{\bk},\widehat{\bq})
+ {3 \over 5}R_{04}(k, b_1) R_{04}(q, b_2)
Y_{(44)0}(\widehat{\bk},\widehat{\bq})\ , \nonumber \\
\label{eq18}
\end{eqnarray}
where $b_1=1/4\gamma$ and $b_2=3/16\gamma$ with $\gamma=\mu\nu=2\nu$,
$R_{n\ell}(x,\nu)$, the radial part of the h.o. wave function,
and $Y_{(\lambda \ell)L}(\widehat{\bk},\widehat{\bq})$ the coupled
angular-momentum functions.
The difference between the Faddeev calculation
and the variational calculation is very small.
In particular the difference of the $3\alpha$ energies
is less than 1 keV both in the Coulomb on and off cases.

In summary, we have found that the Faddeev equation using 2-cluster
RGM kernel, derived in the previous paper, \cite{FU01} may
involve redundant components, if multi Pauli-forbidden states
exist for the 2-cluster relative motion as in the $3\alpha$ system.
For systems of three identical bosons, these components
do not contribute to the total wave function, since they
belong to [21] symmetry with respect to the permutations of 
the three bosons.
We have proposed a simple method to eliminate these redundant components
through diagonalization procedure of the rearrangement matrix
in the 3-cluster model space with the 2-cluster forbidden components.
The equivalence between this modified  Faddeev equation and
the variational approach using the translationally invariant
harmonic-oscillator basis is numerically shown for the ground state
of the $3\alpha$ system.


This work was supported by a  Grant-in-Aid for Scientific
Research from the Ministry of Education, Science, Sports and
Culture (No. 12640265).




\begin{table}[ht]
\caption{The $L=0$ lowest eigen-values
for $2\alpha$ and $3\alpha$ systems,
obtained by diagonalization using [3] symmetric
translationally-invariant h.o. basis. 
$N$ stands for the maximum total h.o. quanta
included in the calculation,
and $n_{\rm max}$ the number of the basis states.
The numbers without the parentheses indicate the results
without the Coulomb force, while those in the parentheses
when the shielded Coulomb force with $R_C=6~{\rm fm}^{-1}$
is included.}
\label{table4}
\begin{center}
\renewcommand{\arraystretch}{1.2}
\setlength{\tabcolsep}{1mm}
\begin{tabular}{cccccc}
\hline
$N$ & $n_{\rm max}$ & $E(2\alpha)$ & $\varepsilon$
& $E(3\alpha)$ & $c_{(04)}$ \\
\hline
4  & $-$ & 20.7705 (23.1969) & $-$  & $-$ & $-$ \\
6  & $-$ &  9.7474 (1.8262) & $-$  & $-$ & $-$ \\
8  &   1 &  5.0088 (6.9205) & 22.7245 (25.1180) & 5.4467 (12.6290)
   & 1 \\
10 &   3 &  2.5254 (4.3239) & 15.9632 (18.0165) & $-2.9705$ (3.6021)
   & 0.9567 (0.9528) \\
12 &   7 &  1.1182 (2.8321) & 12.9965 (14.8448) & $-6.6379$ ($-0.3657$)
   & 0.9160 (0.9077) \\
14 &  12 &  0.2649 (1.9122) & 11.3357 (13.0258) & $-8.5469$ ($-2.4866$)
   & 0.8860 (0.8736) \\
16 &  19 & $-0.2787$ (1.3140) & 10.3336 (11.8894) & $-9.6227$ ($-3.7228$)
   & 0.850 (0.8490) \\
18 &  28 & $-0.6396$ (0.9073) & 9.6987 (11.1403) & $-10.2698$ ($-4.4949$)
   & 0.8511 (0.8315) \\
20 &  39 & $-0.6676$ (0.6203) & 9.2852 (10.6295) & $-10.6721$ ($-4.9962$)
   & 0.8415 (0.8190) \\
22 &  52 & $-1.0629$ (0.4111) & 9.0119 (10.2742) & $-10.9281$ ($-5.3310$)
   & 0.8349 (0.8101) \\
24 &  68 & $-1.1899$ (0.2544) & 8.8295 (10.0235) & $-11.0934$ ($-5.5590$)
   & 0.8305 (0.8037) \\
26 &  86 & $-1.2839$ (0.1343) & 8.7071 (9.8446) & $-11.2014$ ($-5.7168$)
   & 0.8275 (0.7990) \\
28 & 107 & $-1.3547$ (0.0403) & 8.6246 (9.7157) & $-11.2726$ ($-5.8275$)
   & 0.8255 (0.7957) \\
30 & 131 & $-1.4090$ ($-0.0346$) & 8.5686 (9.6222) & $-11.3198$ ($-5.9061$)
   & 0.8240 (0.7932) \\
32 & 158 & $-1.4511$ ($-0.0952$) & 8.5306 (9.5538) & $-11.3514$ ($-5.9624$)
   & 0.8232 (0.7914) \\
34 & 188 & $-1.4842$ ($-0.1448$) & 8.5047 (9.5034) & $-11.3727$ ($-6.0032$)
   & 0.8225 (0.7901) \\
36 & 222 & $-1.5105$ ($-0.1860$) & 8.4869 (9.4660) & $-11.3870$ ($-6.0329$)
   & 0.8221 (0.7891) \\
38 & 259 & $-1.5317$ ($-0.2205$) & 8.4746 (9.4381) & $-11.3968$ ($-6.0548$)
   & 0.8218 (0.7884) \\
40 & 300 & $-1.5487$ ($-0.2496$) & 8.4661 (9.4172) & $-11.4035$ ($-6.0711$)
   & 0.8216 (0.7878) \\
42 & 345 & $-1.5626$ ($-0.2745$) & 8.4603 (9.4014) & $-11.4081$ ($-6.0831$)
   & 0.8214 (0.7874) \\
44 & 394 & $-1.5741$ ($-0.2959$) & 8.4562 (9.3895) & $-11.4113$ ($-6.0922$)
   & 0.8213 (0.7871) \\
46 & 447 & $-1.5835$ ($-0.3144$) & 8.4533 (9.3804) & $-11.4135$ ($-6.0991$)
   & 0.8213 (0.7869) \\
48 & 505 & $-1.5914$ ($-0.3305$) & 8.4513 (9.3734) & $-11.4150$ ($-6.1043$)
   & 0.8212 (0.7867) \\
50 & 567 & $-1.5979$ ($-0.3446$) & 8.4499 (9.3680) & $-11.4161$ ($-6.1082$)
   & 0.8212 (0.7865) \\
52 & 634 & $-1.6034$ ($-0.3570$) & 8.4489 (9.3639) & $-11.4168$ ($-6.1113$)
   & 0.8211 (0.7864) \\
54 & 706 & $-1.6081$ ($-0.3680$) & 8.4482 (9.3607) & $-11.4174$ ($-6.1136$)
   & 0.8211 (0.7863) \\
56 & 783 & $-1.6120$ ($-0.3778$) & 8.4476 (9.3582) & $-11.4178$ ($-6.1154$)
   & 0.8211 (0.7863) \\
58 & 865 & $-1.6153$ ($-0.3865$) & 8.4473 (9.3563) & $-11.4180$ ($-6.1168$)
   & 0.8211 (0.7862) \\
60 & 953 & $-1.6182$ ($-0.3942$) & 8.4470 (9.3547) & $-11.4182$ ($-6.1179$)
   & 0.8211 (0.7862) \\
62 & 1,042 & $-1.6206$ ($-0.4013$) & 8.4468 (9.3535) & $-11.4184$ ($-6.1188$)
   & 0.8211 (0.7862) \\
64 & 1,145 & $-1.6227$ ($-0.4076$) & 8.4467 (9.3526) & $-11.4185$ ($-6.1194$)
   & 0.8211 (0.7861) \\
66 & 1,250 & $-1.6245$ ($-0.4133$) & 8.4466 (9.3519) & $-11.4186$ ($-6.1200$)
   & 0.8211 (0.7861) \\
68 & 1,361 & $-1.6261$ ($-0.4185$) & 8.4465 (9.3513) & $-11.4186$ ($-6.1204$)
   & 0.8211 (0.7861) \\
70 & 1,478 & $-1.6274$ ($-0.4231$) & 8.4464 (9.3508) & $-11.4186$ ($-6.1207$)
   & 0.8211 (0.7861) \\
72 & 1,602 & $-1.6286$ ($-0.4270$) & 8.4464 (9.3504) & $-11.4187$ ($-6.1210$)
   & 0.8211 (0.7861) \\
\hline
\multicolumn{3}{c}{Faddeev} & 8.4465 (9.3490) & $-11.4187$ ($-6.1217$)
   & 0.8211 (0.7860) \\
\hline
\end{tabular}
\end{center}
\end{table}

\end {document}